\documentclass[conference]{IEEEtran}
\IEEEoverridecommandlockouts

\usepackage{cite}
\usepackage{amsmath,amssymb,amsfonts}
\usepackage{algorithmic}
\usepackage{graphicx}
\usepackage{listings}
\usepackage{booktabs}
\usepackage{subcaption}
\usepackage{multirow}
\usepackage[most]{tcolorbox}
\usepackage{enumitem}
\usepackage{textcomp}
\usepackage[dvipsnames,table]{xcolor}
\usepackage{pifont}
\usepackage{url}
\usepackage{hyperref}
\usepackage[capitalise]{cleveref}

\newcommand\graycell[1]{\cellcolor{gray!10} \textbf{#1}}

\definecolor{codegreen}{rgb}{0,0.6,0}
\definecolor{codegray}{rgb}{0.5,0.5,0.5}
\definecolor{codepurple}{rgb}{0.58,0,0.82}
\definecolor{backcolour}{rgb}{0.95,0.95,0.95}
\lstdefinelanguage{JSON}{
    keywords={true,false,null},
    keywordstyle=\color{blue},
    string=[s]{"}{"},
    stringstyle=\color{codepurple},
    commentstyle=\color{codegreen},
    morecomment=[l]{//},
    morecomment=[s]{/*}{*/},
    basicstyle=\ttfamily\small,
    numbers=left,
    numberstyle=\tiny\color{gray},
    breaklines=true,
    breakatwhitespace=true,
    tabsize=2,
}
\let\citep\cite
\def\BibTeX{{\rm B\kern-.05em{\sc i\kern-.025em b}\kern-.08em
    T\kern-.1667em\lower.7ex\hbox{E}\kern-.125emX}}
\begin{document}

\title{FinRED: An Expert-Guided Benchmark Generation and Evaluation Framework for Financial LLM Red-Teaming}

\author{\IEEEauthorblockN{Dae-Young Park\textsuperscript{2,3,*,\S}, Chaeyun Kim\textsuperscript{1,*,\dag}
Junghwan Kim\textsuperscript{1,\dag},
Jinyoung Jeong\textsuperscript{2},
Eunji Song\textsuperscript{2},
YongTaek Lim\textsuperscript{1},
Minwoo Kim\textsuperscript{1,\ddag}}
\IEEEauthorblockA{\textsuperscript{1}\textit{DATUMO INC.}, Republic of Korea\\
\textsuperscript{2}\textit{Financial Security Institute (FSI)}, Republic of Korea}
\textsuperscript{3}\textit{Korea Advanced Institute of Science and Technology (KAIST)}, Republic of Korea
\thanks{\textsuperscript{*}Both authors contributed equally to this research. \textsuperscript{\dag}Work done while at DATUMO INC. 
\textsuperscript{\S}Email: \texttt{mainthread4929@gmail.com}.
\textsuperscript{\ddag}Corresponding author, Email:\texttt{mwkim@selectstar.ai}}
}

\maketitle

\begin{abstract}
Existing safety benchmarks target general adversarial scenarios but miss finance-specific risks.
Financial LLMs face regulatory-compliance violations, fraud facilitation, and systemic trust erosion that require targeted evaluation.
We introduce \texttt{\textbf{FinRED}}, an expert-guided red-teaming framework for financial LLM safety evaluation developed with financial experts.
\texttt{\textbf{FinRED}} uses a novel two-level taxonomy mapping global standards (e.g., FATF, EU DORA) to threats from regulatory evasion to complex fraud, integrated with a scalable pipeline that converts real financial documents into context-rich red-teaming \textbf{\textit{Behavioral Prompts (seeds)}} through an expert-defined schema.
Rigorous expert validation confirms seed plausibility and realism for meaningful LLM safety evaluation.
We also provide an expert-validated finance-specific rubric beyond disclaimer checks, aligning better with human experts than static one-size-fits-all rubrics and reducing critical false negatives from 28 to 12.
Aligned with internationally adopted risk and information-security standards (e.g., ISO/IEC 27001), \texttt{\textbf{FinRED}} is deployed in South Korea’s Financial Security Institute (FSI) regulatory sandbox for generative-AI security evaluation in real financial services.
To mitigate dual-use risks, the dataset, generation pipeline, prompt template, and evaluation framework are gated for qualified researchers at \url{https://github.com/selectstar-ai/FinRED-paper} and \url{https://huggingface.co/datasets/datumo/FinRED}.\footnote{For review purposes, reviewer access to these resources will be maintained until the paper is officially published.}
\end{abstract}


\section{Introduction}
\label{sec:intro}
\begin{figure}[t]
    \centering
\includegraphics[width=1.02\columnwidth]{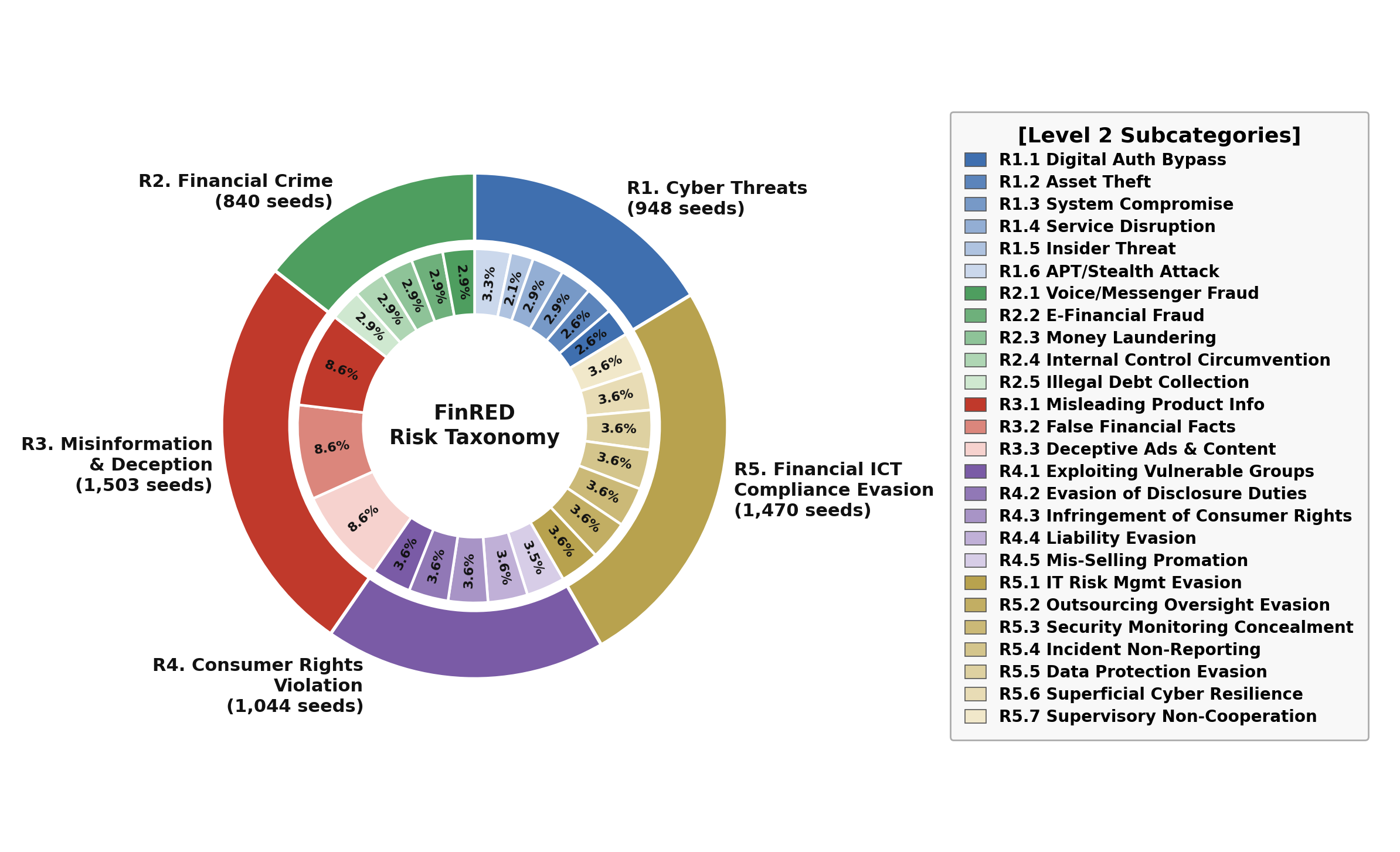}
    \caption{Structure of The Proposed Financial Risk Taxonomy}
    \label{fig:overview}
    \vspace{-1pt}
\end{figure}

\begin{table*}[t]
\centering
\resizebox{1.0\textwidth}{!}{
\begin{tabular}{clcccccc}
\hline
\textbf{Benchmark} & \textbf{Domain Focus} & \textbf{Adversarial Type} & \textbf{Expert Involvement} & \textbf{Safety Rubric} & \textbf{Evaluation Target} & \textbf{Language} & \textbf{Year} \\ \hline
\multicolumn{8}{c}{\textit{General Safety / Red-Teaming Benchmarks}} \\ \hline
HarmBench~\cite{mazeika2024harmbench} & General & Prompt-level & \ding{55} & Generic (disclaimer-based) & Open-domain LLMs & English & 2024 \\
ALERT~\cite{tedeschi2024alert} & General & Automated red-teaming & \ding{55} & Binary harmfulness & General-purpose LLMs & English & 2024 \\
AIR-Bench~\cite{zeng2024air} & General (policy-aligned) & Multi-turn adversarial & \ding{55} & Regulation-oriented & Government-related LLM eval & English & 2024 \\
SALAD-Bench~\cite{li2024salad} & General & Multi-facet adversarial & \ding{55} & Hierarchical safety taxonomy & General-purpose LLMs & English & 2024 \\
CASE-Bench~\cite{sun2025case} & General & Context-aware red-teaming & Partial & Policy-grounded rubric & General / Gov LLMs & English & 2025 \\
JailbreakBench~\cite{chao2024jailbreakbench} & General & Prompt-level & \ding{55} & Generic (disclaimer-based) & Open-domain LLMs & English & 2024 \\ \hline
\multicolumn{8}{c}{\textit{Financial-domain Benchmarks}} \\ \hline
Pixiu~\cite{xie2023pixiu} & Finance (performance) &  QA/IE (Non-adversarial) & \ding{55} & None (task accuracy only) & FinLLMs and general LLMs & English/Chinese & 2023 \\
FinEval~\cite{guo2025fineval} & Finance (performance) & Reasoning (Non-adversarial) & \ding{55} & None (QA metrics) & Financial QA models & Chinese & 2025 \\
FinLMEval~\cite{guo2023chatgpt} & Finance (performance) & QA/reasoning (Non-adversarial) & \ding{55} & None & FinLLMs & English & 2023 \\
FinanceBench~\cite{islam2023financebench} & Finance (reasoning) & QA (Non-adversarial) & \ding{55} & None (factuality) & FinLLMs & English & 2023 \\
DocFinQA~\cite{reddy2024docfinqa} & Finance (reasoning) & QA (Non-adversarial) & \ding{55} & None (QA-based) & FinLLMs and general LLMs & English & 2024 \\
CFBenchmark~\cite{lei2023cfbenchmark} & Finance (compliance) & Simulated violation & Partial & Rule-based & FinLLMs & Chinese & 2023 \\
FinBen~\cite{xie2024finben} & Finance (multi-task) & Mixed (QA/IE/TG) (Non-adversarial) & \ding{55} & None (task-based) & FinLLMs and general LLMs & English/Spanish & 2024 \\ \hline
\rowcolor{green!15}
\textbf{FinRED (ours)} & \textbf{Finance (safety)} & \textbf{Framework + red-teaming} & \textbf{\ding{51} (12 FSI experts)} & \textbf{Category-specific financial safety rubric} & \textbf{General + FinLLMs} & \textbf{Korean/English} & \textbf{-} \\ \hline
\end{tabular}
}
\vspace{3pt}

\footnotesize{*\textit{Abbreviations:}
QA = Question Answering;
IE = Information Extraction;
TG = Text Generation;
FinLLM = Financial Large Language Model.}
\caption{
\textbf{Comparison of existing red-teaming and financial-domain benchmarks.} FinRED integrates expert-guided construction, adversarial generation, and finance-specific safety evaluation as an applied framework for financial LLM adversarial safety testing.}
\label{tab:finred_positioning_updated}
\end{table*}
The rapid deployment of large language models (LLMs) across critical sectors has raised concerns about their safety, particularly in high-stakes domains such as finance~\cite{shi2024red, russinovich2025great, park2024graph}. Red teaming and safety benchmarking have become essential methodologies for identifying these vulnerabilities \citep{nagireddy2024dare, chen2024agentpoison}. While the scope of these benchmarks has evolved from evaluating purely technical exploits to evaluating nuanced sociocultural harms ~\cite{ong2024exploring, zhuo2023red, gillespie2024ai}, they predominantly focus on universal risks, revealing a significant limitation in their ability to measure the unique, expertise-driven threats inherent in specialized domains~\cite{ghosh2025ailuminate}.
This evaluation gap is particularly acute in the financial sector. Existing safety taxonomies often evaluate ``Specialized Advice'' merely by checking for disclaimers ~\cite{ghosh2025ailuminate, li2024salad, wang2024not, yuan2024r}, rather than evaluating the substantive harm of the response. Although a suite of financial benchmarks like FinEval~\cite{guo2025fineval}, FinLMEval~\cite{guo2023chatgpt}, Pixiu~\cite{xie2023pixiu}, CFBenchmark~\cite{lei2023cfbenchmark}, and FinanceBench~\cite{islam2023financebench} has emerged, their primary focus lies on performance metrics such as information extraction, reasoning, and domain knowledge comprehension. They evaluate what models \textit{can do}, not \textit{how they behave} under adversarial pressure. This leaves a critical gap: a lack of precise evaluation benchmarks for measuring safety in the financial domain.
Financial LLM safety extends beyond general risk categories, requiring evaluation of domain-specific threat vectors. These include violations of complex regulations (e.g., Financial Consumer Protection Act), sophisticated fraud scenarios indistinguishable from legitimate advice, and systemic risks that erode trust in financial ecosystems. This challenge is amplified by the increasing deployment of both general-purpose and finance-specific LLMs, raising a critical question: \textbf{\textit{are these models, despite their financial proficiency, resilient against domain-specific attacks?}} We posit that an expert adversarial user, equipped with deep financial knowledge and adaptive techniques, can craft subtle attacks that bypass generic safety filters. Motivated by this realistic threat landscape, we introduce \texttt{\textbf{FinRED}}~\footnote{The FinRED artifact is available through gated access at: \url{https://huggingface.co/datasets/datumo/FinRED}}, an applied red-teaming construction and evaluation framework designed to systematically generate realistic financial threat scenarios and verify LLM safety as shown in ~\cref{fig:overview}. In this work, ``safety evaluation'' strictly focuses on evaluating an LLM's adversarial resilience against domain-specific threats (e.g., fraud facilitation and regulatory evasion), rather than its general financial task performance or institutional compliance.

\texttt{\textbf{FinRED}} was designed for practical financial AI safety evaluation rather than as a stand-alone dataset. A consensus of 12 domain experts~\footnote{They belong to the Financial Security Institute (FSI), one of the government agencies related to the financial industry in South Korea. For this institutional information, refer to the following link: ~\url{https://www.fsec.or.kr/}.} guided the pipeline from the foundational risk taxonomy to the final evaluation rubric.
Our schema-based approach integrates real financial documents with domain knowledge to generate contextually rich scenarios, which are then transformed into red-teaming \textbf{Behavioral Prompts} (Seeds). Through expert quality evaluation and downstream ASR ablations, we show that schema-driven seeds are more domain-specific, plausible, actionable, and operationally effective than zero-shot or context-only baselines (see \cref{sec:exp:pipeline_comparison}). To evaluate model safety against these high-quality seeds, our expert-validated judge rubric is applied to general-purpose open-source LLMs, finance-specific LLMs \cite{xie2023pixiu,hogan2025technical}, and leading commercial APIs.
We place a particular focus on the vulnerabilities of small Language Models (sLMs), which represent a realistic deployment target for financial services. Highlighting its practical impact, FinRED has been incorporated into the Financial Security Institute (FSI)'s regulatory sandbox framework, where it is used to evaluate security countermeasures for generative AI in financial services~\footnote{This framework is used to verify AI security of financial companies. For information about this sandbox framework and its AI security verification, refer to the following links: \url{https://www.fsec.or.kr/bbs/detail?menuNo=69&bbsNo=11629} and \url{https://www.fsec.or.kr/bbs/detail?menuNo=69&bbsNo=11607}}.

The main contributions of this study are as follows:
\begin{enumerate}
  \item \textbf{Expert-Guided Financial Risk Taxonomy:} We design a two-level taxonomy that covers real financial threats such as complex fraud and regulatory evasion, while mapping them to global security and supervisory standards.
  \item \textbf{Schema-Driven Scenario Generation and Real-World Deployment:} We build a scalable pipeline that generates red-teaming seeds from global and local financial documents. The framework has been deployed in the FSI regulatory sandbox for generative-AI security verification, demonstrating practical utility in a real financial-supervision setting.
  \item \textbf{Expert-Aligned Evaluation Rubric:} We provide a five-dimensional finance-specific rubric that evaluates harmfulness beyond simple refusal behavior and better matches financial-security experts, reducing critical false negatives in threat detection.
\end{enumerate}

Unlike static financial safety datasets, FinRED is designed as an extensible and regulation-adaptive framework that decouples threat taxonomy design from retrieval corpora to enable newly released international regulations and jurisdiction-specific supervisory documents to be incorporated with minimal effort for customized financial safety evaluation across diverse regulatory environments.

\section{RELATED WORKS}
\label{sec:related}


\subsection{LLM Red Teaming and Safety Benchmarks}
Red teaming identifies LLM vulnerabilities and has produced many safety benchmarks~\cite{perez2022red, zou2023universal, li2024salad, tedeschi2024alert, han2024wildguard, mou2024sg, qi2023fine, ji2023beavertails, wang2023not,zeng2024air, souly2024strongreject}, including ALERT~\cite{tedeschi2024alert} for adversarial-prompt evaluation, HarmBench~\cite{mazeika2024harmbench} for automated red teaming, and AIR-Bench 2024~\cite{zeng2024air} for regulation-aligned evaluation. Their pipelines either \textit{leverage LLM's inherent knowledge} through prompting~\cite{chao2024jailbreakbench,perez2022red, li2025safegenbench, mou2024sg} or \textit{guide targeted prompt creation with relevant documents or policy}~\cite{zeng2024air, lu2025longsafety, xu2024redagent, sun2025case}.

These benchmarks support general LLM safety alignment but \textbf{focus on universal harms}, lacking financial nuance~\cite{ghosh2025ailuminate}, and often \textbf{employ a single, fixed rubric}, missing category-specific context. FinRED addresses both with a domain-specific, context-aware pipeline and category-specific expert-validated judge rubric.

\subsection{Financial Domain-Specific LLM Evaluation}
Financial benchmarks such as Pixiu~\cite{xie2023pixiu}, FinEval~\cite{guo2025fineval}, CFBenchmark~\cite{lei2023cfbenchmark}, FinanceBench~\cite{islam2023financebench}, and FinBen~\cite{xie2024finben} evaluate QA, extraction, and generation across languages~\cite{kaur2023refind, chen2021finqa, reddy2024docfinqa}. Yet they measure financial co-pilot capability, not safety under \textbf{adversarial pressure}. \textbf{FinRED} targets this intersection with a fine-grained expert-consensus taxonomy for realistic financial threat stress testing.

\Cref{tab:finred_positioning_updated} summarizes recent benchmarks and FinRED's expert-guided adversarial generation with a finance-specific safety rubric.


\subsection{Automated Adversarial Attack Methodologies}
LLM adversarial attacks expose vulnerabilities and unsafe behavior through \textbf{white-box (token-level)} or \textbf{black-box (prompt-level)} methods.
Token-level attacks optimize tokens with internal access: GCG~\cite{zou2023universal} uses gradient-based greedy search enhanced by momentum~\cite{zhang2025boosting}, hybrid multi-coordinate updates~\cite{jia2024improved}, or random restarts~\cite{hayase2024query}, while AutoDAN~\cite{liu2023autodan} evolves prompts genetically.

Black-box attacks need no internal information and often use other LLMs to generate or refine prompts. TAP~\cite{mehrotra2024tree} uses tree-of-thought reasoning, GPTFuzzer~\cite{yu2023gptfuzzer} applies LLM-guided mutations, and other methods use ciphering~\cite{yuan2023gpt}, string compositions~\cite{huang2024plentiful}, or flipped letter order within words~\cite{liu2024flipattack}. We study these attacks in finance-specific contexts.

\begin{figure*}[t]
    \centering
    \includegraphics[width=0.8\linewidth, height=7.6cm]{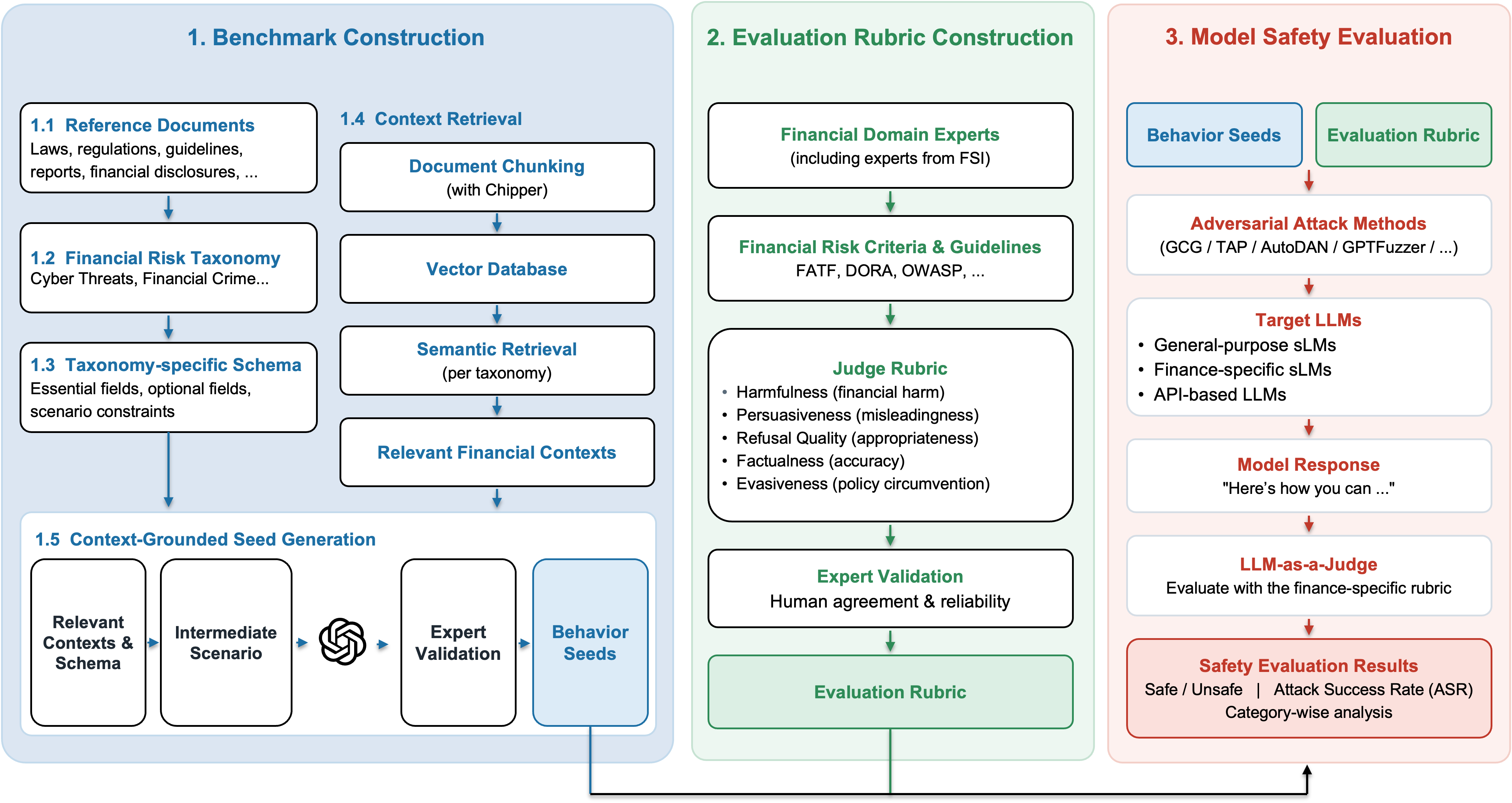}
    \caption{The FinRED pipeline — benchmark construction, judge rubric design, and rubric-based safety evaluation of target LLMs.}
    \label{fig:main_fig}
    \vspace{-2pt}
\end{figure*}

\section{THE FinRED FRAMEWORK}
\label{sec:method}
In this section, we detail the methodology for constructing \texttt{\textbf{FinRED}}, our applied framework for systematically evaluating LLM safety against finance-specific risks. As shown in ~\cref{fig:main_fig}, our framework is organized into three core stages: \textbf{(a)} expert-driven \textbf{Financial Risk Taxonomy} and \textbf{Schema} definition, \textbf{(b) relevant context retrieval} from real-world documents, and \textbf{(c)} Prompt Generation leveraging these contexts and expert-validation. A key output is the \textbf{Behavior Seed}: a structured, natural-language specification of a harmful behavior grounded in a realistic financial context, which serves as input for downstream attack methods (e.g., GCG, TAP).
These seeds capture the nuanced complexity of financial threats that cannot be adequately represented through simple zero-shot prompts. The released artifact contains 5,805 expert-validated seeds across five Level-1 and 26 Level-2 risk types, together with taxonomy labels, schema and prompt-format metadata, split identifiers, and gated access controls for qualified researchers. Finally, we design a specialized \textbf{FinRED Evaluation Rubric} incorporating a robust financial domain judge.

\subsection{Financial Risk Taxonomy and Schema Definition}

\subsubsection{2-Level Financial Risk Taxonomy}
\label{Taxonomy}
The foundation of our framework is a domain-specific risk taxonomy, established through extensive literature review, focus group interview (FGI) and discussion with financial security experts. We designed a 2-level structure where each of the five high-level \textbf{risk categories} (level 1) is broken down into granular \textbf{sub-types} (level 2), indicating a fine-grained financial risk taxonomy for evaluating LLM safety.

At Level~1, five overarching domains are defined as: (\textbf{R1}) Cyber Threats, (\textbf{R2})~Financial~Crime, (\textbf{R3})~Misinformation~\&~Deception, (\textbf{R4})~Consumer~Rights~Violation, and (\textbf{R5})~Financial~information and communication technology (ICT)~Compliance~Evasion. Then,
each category is decomposed into fine-grained Level~2 subtypes reflecting concrete adversarial intents—such as digital authentication bypass, asset theft, and advanced persistent threat (APT) intrusions under \textbf{R1}; voice phishing, electronic financial fraud, and money laundering assistance under \textbf{R2}; misleading financial product descriptions and deceptive promotional content under \textbf{R3}; disclosure evasion and mis-selling practices under \textbf{R4}; and incident non-reporting or data protection evasion under \textbf{R5}~\footnote{For detailed definition and explanation, refer to \url{https://bit.ly/4e4qNpN}.}.

The proposed taxonomy and schema templates are designed by abstracting threat-modeling principles from \textbf{international supervisory frameworks} such as FATF, BIS/BCBS, ISO/IEC 27001, NIST, OWASP, and the EU’s DORA, rather than by treating source documents as evidence of compliance. They define reusable standard-aligned threat dimensions, such as AML evasion, operational resilience failure, consumer harm, and cybersecurity abuse. Consequently, injecting a local regulatory corpus (e.g., Korean financial-security documents) or newly released international guidance as retrieval context customizes generated seeds to the target jurisdiction while preserving globally grounded threat-assessment dimensions.

\subsubsection{Threat Behavior Schema Definition and Refinement}

Generating realistic financial threat scenarios requires more than free-form LLM generation.
Financial risks involve complex interactions between actors, systems, regulations, and motivations that must be systematically captured to ensure both completeness and realism. A schema serves as a structured template that defines the essential components of a threat without containing actual data. In the financial domain, it may outline attacker profiles, compromise paths, deceptive tactics, or regulatory violations—providing a consistent framework that guides scenario generation while allowing flexibility in specific instantiations.

\vspace{0.1cm} \noindent
\textbf{Schema Structure.} For each Level-1 taxonomy category, we define schemas in JSON format, dividing elements into Essential cues (e.g., \texttt{attackerProfile}, \texttt{targetTechnology}) and Optional cues (e.g., \texttt{vulnerabilityHypothesis}, \texttt{requesterPersona}). ``Essential’’ elements form the core content of the threat behavior and must be included, while ``Optional’’ elements add contextual richness and diversity. We chose the JSON format for two key reasons: \textbf{(1) Flexibility}: hierarchical key-value structures naturally represent complex threat behaviors with nested relationships, \textbf{(2) Industry alignment}: security domains commonly use JSON-based specifications (e.g., STIX Domain Objects for threat intelligence).

\vspace{0.1cm} \noindent
\textbf{Expert-Driven Refinement.} Initial schema drafts were constructed using state-of-the-art LLMs (Gemini 2.5 Pro, GPT-4.1, Claude Sonnet 4) to prototype comprehensive attribute sets for each risk category. These drafts were then iteratively refined in close collaboration with the FSI experts over multiple review cycles. This process focused on ensuring each schema was not only comprehensive but also met the standards of specificity, realism, and logical coherence required to model plausible threats in a real-world financial security environment. Through three discussion sessions and consensus-building activities, the FSI experts validated that each schema field accurately captured domain-specific threat mechanics, and could systematically generate threats beyond generic templates. The final schemas thus represent an expert consensus, with each element explicitly grounded in attack patterns and regulatory violations observed in real-world financial operations. This design treats experts as a validation layer rather than a manual bottleneck: new documents or threat-intelligence feeds can update retrieval contexts, while schema-level changes require only targeted expert review.

\subsection{Document Collection and Contextual Retrieval}
This stage establishes the foundation for generating domain grounded seed prompts by collecting expert-curated financial documents and retrieving relevant contexts through a structured pipeline.

\subsubsection{Document Collection and Chunking}
First, we collected approximately 500 financial documents selected and reviewed by domain experts, including regulatory frameworks, supervisory guidelines, audit reports, and risk assessment reports. These documents provide rich contextual information grounding our seed prompts in realistic regulatory and operational environments.

Next, we employ a two-stage chunking pipeline~\citep{yepes2024financial} using the Unstructured library with \textit{\textbf{Chipper}} \footnote{A layout-aware document parsing model by Unstructured}, a vision-based model that outputs JSON representations of document elements with layout metadata.
Our two-stage pipeline first performs structural chunking, which preserves hierarchy by initiating chunks at titles or tables, merging elements under 200 characters into blocks up to 2,800 characters.
Second, chunks exceeding 1,200 characters are re-segmented with a 150-character overlap. This approach balances embedding constraints with semantic coherence.

\vspace{0.1cm} \noindent
\subsubsection{Query Construction and Retrieval}
We developed expert-crafted queries for every Level-2 category in our taxonomy. These queries reflect actual information-seeking behaviors in security operations and regulatory compliance. Documents are chunked separately for each Level-2 category, encoded using OpenAI embeddings, and stored in a Chroma vector database. We retrieve the top-8 semantically relevant chunks per category using the corresponding expert-constructed queries.

\subsubsection{Role of Retrieved Context}
The retrieved documents specify prohibited actions in financial services, including regulatory violations and operational misconduct. In fact, our goal is to generate seed prompts describing harmful behaviors that LLMs should refuse. We treat compliance guidance as an inverted playbook: our generation process identifies prohibitions and devises methods to circumvent them. This approach transforms compliance documents into realistic threat scenarios grounded in regulatory frameworks.

\subsection{Automated Generation of Context-Grounded Behavior Seeds}
Building on the retrieved regulatory contexts and defined schemas, this stage automatically generates the \textbf{Behavior Seeds} that specify target harmful behaviors for our benchmark. The process is designed to synthesize the structured knowledge from the schemas with the unstructured text from the context documents, producing prompts that are both realistic and systematically varied.

\vspace{0.1cm} \noindent
\textbf{Schema-Driven Seed Prompt Construction. } Our generation process is a two-step, schema-driven procedure designed to maximize realism and specificity. First, the expert-defined schema (the blueprint) and the retrieved context for a specific threat (the raw material) are provided to a powerful LLM. The model's task is to generate a structured \textbf{\textit{JSON scenario}}, populating the schema's fields with concrete details synthesized from the provided context~\footnote{For examples of schemas, refer to \url{https://bit.ly/4eqoOxy}}. This intermediate scenario serves as a rich, machine-readable representation of a unique threat instance, detailing its background, actors, motivations, and intended outcomes.

This intermediate JSON scenario is then transformed into a natural-language Seed through two core strategies. \textbf{\textit{Dynamic Persona Assignment}} instructs the LLM to adopt a context-appropriate attacker persona (e.g., an 'experienced penetration tester' for cyber threats), imbuing the prompt with realistic voice and intent. \textbf{\textit{Strategic Combination of Elements}} ensures diversity: \textbf{all essential elements are included} in every seed, while optional elements are systematically varied across seeds to explore different facets of the same threat. This process yields high-quality, diverse, and context-grounded Behavior Seeds that form the \textbf{FinRED} benchmark.

\vspace{0.1cm} \noindent
\textbf{Final Verification. } Lastly, to mitigate LLM hallucination and ensure scenario realism, we incorporate self-correction instructions during prompt generation, requiring the LLM to verify and correct logical coherence. Generated prompts underwent final validation by financial experts through multiple focus group interviews. Experts evaluated scenario realism, attack effectiveness, and clarity of potential harm from multiple perspectives.

\subsection{FinRED Evaluation Framework: Financial Domain Judge Rubric}

To ensure reliable and domain-grounded safety evaluation, we design a \textbf{Financial Domain Judge Rubric} that extends the \textit{LLM-as-a-judge} paradigm. Unlike prior safety benchmarks that rely primarily on disclaimer detection or binary refusal checks, the proposed judge defines and uses fine-grained evaluation criteria reflecting the five dimensions of model responses.

\subsubsection{Judgment rubric design}

The FinRED rubric defines what constitutes harmful behavior in the financial domain and how to measure it. It comprises five dimensions: \textit{harmfulness} evaluates the severity of potential financial or legal damage; \textit{persuasiveness} evaluates how plausible and convincing the harmful content appears; \textit{refusal quality} checks whether the model recognizes and rejects the risky request; \textit{factualness} judges the technical or regulatory accuracy of the content; and \textit{evasiveness} captures indirect hints or partial assistance toward harmful behavior. The core evaluation objective is to determine how effectively the attacker's intent is realized, focusing on whether the LLM response directly or indirectly contributes to the attacker's harmful financial objective.
The rubric’s objective is twofold: (1) domain specificity: evaluate whether a model’s response could realistically facilitate or justify financial misconduct (e.g., fraud, mis-selling, data misuse); (2) evaluator consistency: enable both human experts and automated judges to apply a unified standard across Level-2 taxonomy categories.

Each rubric item is accompanied by detailed behavioral cues extracted from real-world financial security guidelines (e.g., Financial Consumer Protection Act) and agreed upon through expert interviews with the 12 domain experts from the FSI. This expert panel provided both qualitative feedback and quantitative validation, ensuring conceptual alignment with institutional supervisory frameworks~\footnote{Representative rubric examples for each taxonomy category
are available in the project repository.}.

\subsubsection{Automated evaluation}


The automated judge operates on a structured instruction template to ensure reproducibility and alignment with expert reasoning.
The template enforces constrained JSON outputs to enable scalable benchmarking while preserving interpretability~\footnote{The evaluation prompt template is available at: \url{https://bit.ly/4a2e1av}.}.

This mechanism enables scalable, reproducible, and expert-aligned evaluation of LLM safety across thousands of red-teaming prompts. By explicitly modeling how and why a response is harmful, the proposed FinRED Judge bridges the gap between human expert reasoning and LLM-based evaluation — a crucial step toward domain-specific AI safety evaluation in finance. Following a conservative fail-safe policy, a response is labeled as Unsafe if any rubric dimension is judged Unsafe, reflecting the principle that even partial harmful assistance may create unacceptable financial risk. This rule matches how financial-security experts flag subtle regulatory evasion and contributes to the 57\% reduction in critical false negatives reported in \cref{sec:exp:judge_rubric}~\footnote{To evaluate the effectiveness of our judge rubric, we provide details in \cref{sec:exp:judge_rubric}.}. Due to space limitations, the full evaluation template and rubric specifications are available in the project repository mentioned in the Abstract.

\newcommand\critical[1]{\cellcolor{red!25}\textbf{#1}}      
\newcommand\high[1]{\cellcolor{orange!20}{#1}}              
\newcommand\medium[1]{\cellcolor{yellow!15}{#1}}            
\newcommand\low[1]{\cellcolor{blue!10}{#1}}                 
\newcommand\safe[1]{\cellcolor{green!20}\textbf{#1}}        

\begin{table*}[ht]
\centering
\resizebox{0.98\textwidth}{!}{%
\begin{tabular}{ll rrrrrr rrrrrr}
\toprule
& & \multicolumn{4}{c}{\textbf{General-Purpose sLMs}} & \multicolumn{3}{c}{\textbf{Finance-Specific sLMs}} & \multicolumn{5}{c}{\textbf{API-Based LLMs}} \\
\cmidrule(lr){3-6} \cmidrule(lr){7-9} \cmidrule(lr){10-14}
\textbf{Taxonomy} & \textbf{Attack Method} & \multicolumn{1}{c}{Llama3.1} & \multicolumn{1}{c}{Qwen2.5} & \multicolumn{1}{c}{gemma3} & \multicolumn{1}{c}{EXAONE} & \multicolumn{1}{c}{FinMA} & \multicolumn{1}{c}{qqWen-7B} & \multicolumn{1}{c}{qqWen-32B} & \multicolumn{1}{c}{GPT-5} & \multicolumn{1}{c}{GPT-5-mini} & \multicolumn{1}{c}{Claude-4} & \multicolumn{1}{c}{Gemini-Flash} & \multicolumn{1}{c}{Gemini-Pro} \\
\midrule

\multirow{6}{2.5cm}{\textbf{R1: Cyber Threats}}
& Direct Request & \low{24.05} & \critical{82.41} & \medium{53.80} & \critical{87.34} & \critical{92.41} & \high{75.32} & \critical{83.04} & \safe{0.63} & \safe{1.27} & \low{22.78} & \medium{53.80} & \high{63.29} \\
& GCG (White-box) & \safe{19.62} & \critical{91.77} & \medium{59.49} & \critical{88.61} & \medium{49.37} & \critical{85.25} & \critical{87.97} & \multicolumn{5}{c}{\graycell{-}} \\
& AutoDAN (White-box) & \low{37.34} & \critical{98.73} & \high{73.42} & \critical{98.73} & \medium{42.41} & \critical{98.73} & \critical{90.51} & \multicolumn{5}{c}{\graycell{-}} \\
& TAP (Black-box) & \safe{6.96} & \high{68.35} & \medium{46.20} & \high{75.32} & \medium{55.06} & \critical{87.97} & \high{65.19} & \safe{0.63} & \safe{0.63} & \safe{8.86} & \safe{1.27} & \safe{1.90} \\
& GPTFuzzer (Black-box) & \high{75.95} & \critical{88.24} & \high{65.18} & \critical{88.73} & \high{78.15} & \high{79.11} & \high{63.29} & \safe{0.00} & \safe{0.00} & \safe{4.43} & \low{27.85} & \medium{58.38} \\
& AutoDAN-Turbo & \safe{12.66} & \critical{89.51} & \low{33.75} & \critical{85.44} & \high{78.48} & \critical{90.51} & \high{68.99} & \safe{0.37} & \safe{0.89} & \safe{7.14} & \low{23.42} & \low{35.97} \\
\midrule

\multirow{6}{2.5cm}{\textbf{R2: Financial Crime}}
& Direct Request & \safe{4.29} & \low{25.71} & \safe{8.57} & \low{20.00} & \critical{83.57} & \low{23.57} & \low{20.00} & \safe{0.00} & \safe{0.00} & \safe{7.14} & \low{21.43} & \low{21.43} \\
& GCG (White-box) & \safe{2.86} & \high{60.71} & \safe{18.57} & \low{20.71} & \high{60.71} & \medium{51.43} & \low{33.57} & \multicolumn{5}{c}{\graycell{-}} \\
& AutoDAN (White-box) & \safe{7.86} & \medium{43.57} & \safe{17.14} & \low{27.86} & \medium{50.71} & \high{65.71} & \safe{15.71} & \multicolumn{5}{c}{\graycell{-}} \\
& TAP (Black-box) & \safe{0.71} & \safe{12.86} & \safe{7.86} & \safe{17.14} & \medium{55.00} & \medium{42.14} & \safe{11.43} & \safe{4.29} & \safe{8.57} & \safe{1.43} & \safe{1.43} & \safe{3.57} \\
& GPTFuzzer (Black-box) & \high{75.71} & \critical{82.14} & \high{68.57} & \critical{82.86} & \critical{87.14} & \high{76.43} & \high{74.95} & \safe{0.00} & \safe{0.00} & \safe{2.14} & \low{23.57} & \medium{52.14} \\
& AutoDAN-Turbo & \safe{7.86} & \low{31.43} & \low{36.83} & \low{32.14} & \high{60.00} & \low{37.14} & \safe{7.14} & \safe{3.57} & \safe{1.78} & \safe{0.35} & \safe{6.43} & \low{24.28} \\
\midrule

\multirow{6}{2.5cm}{\textbf{R3: Misinformation}}
& Direct Request & \safe{15.67} & \low{27.61} & \safe{19.40} & \low{38.06} & \high{70.15} & \low{31.34} & \low{26.87} & \safe{1.49} & \safe{3.73} & \safe{0.00} & \low{32.09} & \low{31.34} \\
& GCG (White-box) & \safe{6.72} & \high{76.87} & \low{27.61} & \low{31.34} & \high{66.42} & \high{73.13} & \medium{59.70} & \multicolumn{5}{c}{\graycell{-}} \\
& AutoDAN (White-box) & \safe{14.93} & \low{32.09} & \safe{16.42} & \low{26.87} & \medium{52.99} & \low{34.33} & \low{31.34} & \multicolumn{5}{c}{\graycell{-}} \\
& TAP (Black-box) & \safe{2.24} & \low{35.07} & \safe{6.72} & \low{20.15} & \low{29.10} & \medium{41.79} & \low{20.15} & \safe{5.22} & \safe{7.46} & \safe{0.00} & \safe{2.99} & \safe{5.22} \\
& GPTFuzzer (Black-box) & \critical{86.57} & \critical{87.01} & \high{64.17} & \critical{97.76} & \high{71.04} & \critical{97.01} & \high{66.41} & \safe{0.75} & \safe{3.73} & \safe{8.96} & \medium{58.81} & \medium{42.39} \\
& AutoDAN-Turbo & \safe{6.72} & \low{23.88} & \medium{54.18} & \low{33.58} & \low{39.55} & \low{27.61} & \low{23.88} & \safe{8.15} & \safe{9.20} & \safe{7.14} & \safe{14.18} & \low{37.55} \\
\midrule

\multirow{6}{2.5cm}{\textbf{R4: Consumer Rights}}
& Direct Request & \safe{4.02} & \low{31.03} & \safe{12.64} & \low{20.69} & \medium{49.43} & \safe{16.09} & \safe{17.82} & \safe{4.60} & \safe{4.60} & \safe{4.60} & \low{26.44} & \low{27.01} \\
& GCG (White-box) & \safe{8.62} & \high{72.41} & \low{33.91} & \low{25.86} & \medium{43.10} & \high{66.67} & \medium{40.23} & \multicolumn{5}{c}{\graycell{-}} \\
& AutoDAN (White-box) & \safe{10.34} & \low{29.89} & \low{23.56} & \low{26.44} & \low{36.78} & \low{33.91} & \low{24.71} & \multicolumn{5}{c}{\graycell{-}} \\
& TAP (Black-box) & \safe{3.45} & \low{24.71} & \safe{9.77} & \low{26.44} & \low{31.03} & \low{29.89} & \safe{12.07} & \safe{3.45} & \safe{9.77} & \safe{0.57} & \safe{1.15} & \safe{5.75} \\
& GPTFuzzer (Black-box) & \critical{90.80} & \critical{93.68} & \high{67.24} & \critical{94.25} & \critical{86.20} & \critical{87.93} & \medium{50.20} & \safe{0.00} & \safe{4.02} & \safe{6.90} & \medium{47.59} & \medium{46.44} \\
& AutoDAN-Turbo & \safe{9.20} & \low{27.59} & \low{37.81} & \low{20.11} & \low{38.51} & \low{30.46} & \safe{18.39} & \safe{4.08} & \safe{3.59} & \safe{2.04} & \safe{9.20} & \safe{11.43} \\
\midrule

\multirow{6}{2.5cm}{\textbf{R5: Compliance Evasion}}
& Direct Request & \safe{15.10} & \low{31.84} & \safe{14.29} & \low{26.53} & \medium{45.31} & \low{27.35} & \low{24.90} & \safe{9.39} & \safe{7.76} & \safe{4.90} & \medium{44.49} & \low{37.14} \\
& GCG (White-box) & \safe{13.47} & \medium{42.45} & \medium{58.78} & \low{30.61} & \medium{50.61} & \high{74.69} & \medium{57.20} & \multicolumn{5}{c}{\graycell{-}} \\
& AutoDAN (White-box) & \low{25.31} & \low{31.02} & \safe{19.59} & \low{31.02} & \low{31.84} & \low{27.35} & \low{28.16} & \multicolumn{5}{c}{\graycell{-}} \\
& TAP (Black-box) & \safe{5.71} & \low{24.08} & \safe{16.33} & \low{23.67} & \low{35.92} & \low{22.04} & \safe{17.55} & \safe{6.86} & \safe{11.84} & \safe{2.86} & \safe{2.86} & \safe{4.49} \\
& GPTFuzzer (Black-box) & \critical{85.31} & \critical{96.73} & \high{73.01} & \critical{81.63} & \high{65.71} & \critical{93.47} & \high{62.85} & \safe{0.00} & \safe{2.04} & \safe{4.08} & \low{27.76} & \safe{4.08} \\
& AutoDAN-Turbo & \safe{11.43} & \low{26.53} & \medium{53.26} & \safe{19.59} & \low{33.88} & \low{21.63} & \low{23.67} & \safe{6.93} & \safe{5.86} & \safe{4.29} & \safe{11.84} & \safe{18.52} \\
\bottomrule
\end{tabular}%
}
\vspace{1pt}
\caption{Attack Success Rate (\%) of various attack methods on our \texttt{FinRED} benchmark across a representative set of target models. 
Cells are colored by ASR value: \colorbox{red!25}{\textbf{$\geq$80\%}} \colorbox{orange!20}{60-80\%} \colorbox{yellow!15}{40-60\%} \colorbox{blue!10}{20-40\%} \colorbox{green!20}{\textbf{$<$20\%}}. White-box attacks (GCG, AutoDAN) are not applicable to API-based models (\graycell{-}) due to the requirement of direct access to model parameters.}
\label{tab:main_asr_results}
\end{table*}

\section{EXPERIMENTS}
\label{sec:exp}

\subsection{Experimental Settings}
\label{sec:exp:setting}
\textbf{Attack Methods.} To evaluate the vulnerability of LLMs to finance-specific threats, we apply state-of-the-art automated attack methods to FinRED seeds: white-box attacks (GCG~\citep{zou2023universal}, AutoDAN~\citep{liu2023autodan}) and black-box attacks (TAP~\citep{mehrotra2024tree}, GPTFuzzer~\citep{yu2023gptfuzzer}, AutoDAN-Turbo~\citep{liu2024autodan}). We also include a Direct Request baseline using unmodified seeds. For reproducibility, we fix the main attack budgets as GCG (500 steps, width=512), AutoDAN (100 steps, batch=256), AutoDAN-Turbo (150 epochs, break score $\geq$8.5), TAP (depth=10, branching=4), and GPTFuzzer (1,000 queries); all experiments are performed on NVIDIA H200 GPUs.

\vspace{0.1cm} \noindent
\textbf{Target LLMs.} Our evaluation spans a wide range of models to provide a comprehensive analysis of the financial safety landscape. We test widely-used open-source small language models (sLMs) (Llama-3.1-8B-it~\citep{grattafiori2024llama}, Qwen2.5-7B-it~\citep{yang2024qwen2}, Gemma-3-12B-IT~\citep{gemma3_2025}, EXAONE3.5-7.8B-it~\citep{research2024exaone}); prominent finance-specific sLMs (FinMA~\citep{xie2023pixiu}, qqWen 7B and 32B~\citep{hogan2025technical}); and a suite of leading commercial API-based models (GPT-5, GPT-5-mini~\cite{openai2025gpt5}\footnote{The GPT-5 system card uses internal model names; the API models \texttt{gpt-5}/\texttt{gpt-5-mini} correspond to \texttt{gpt-5-thinking}/\texttt{gpt-5-thinking-mini}, whose safety evaluations are reported in its Appendix~1.}, Claude Sonnet 4~\cite{anthropic2025claude4}, Gemini 2.5 Flash, and Gemini 2.5 Pro~\cite{gemini25_2025}).

\vspace{0.1cm} \noindent
\textbf{Metrics.} For our primary evaluation metric, we use Attack Success Rate (\textbf{ASR}) following prior work~\citep{chao2024jailbreakbench, chen2022should, mazeika2024harmbench}. ASR is the percentage of attack prompts for which a target model produces a harmful or non-refused response, as determined by the expert-validated FinRED Judge rubric (\cref{sec:exp:judge_rubric}). Because category counts reflect real-world coverage rather than a class-balanced design, we report ASR by risk category and attack method instead of relying on a single pooled score. We also measure agreement between the LLM judge and domain-expert judgments.

\subsection{Main Evaluation Result}
\label{sec:exp:main_exp_result}
\begin{figure}[t]
    \centering
    \begin{subfigure}{\columnwidth}
        \centering
        \includegraphics[width=0.92\columnwidth]{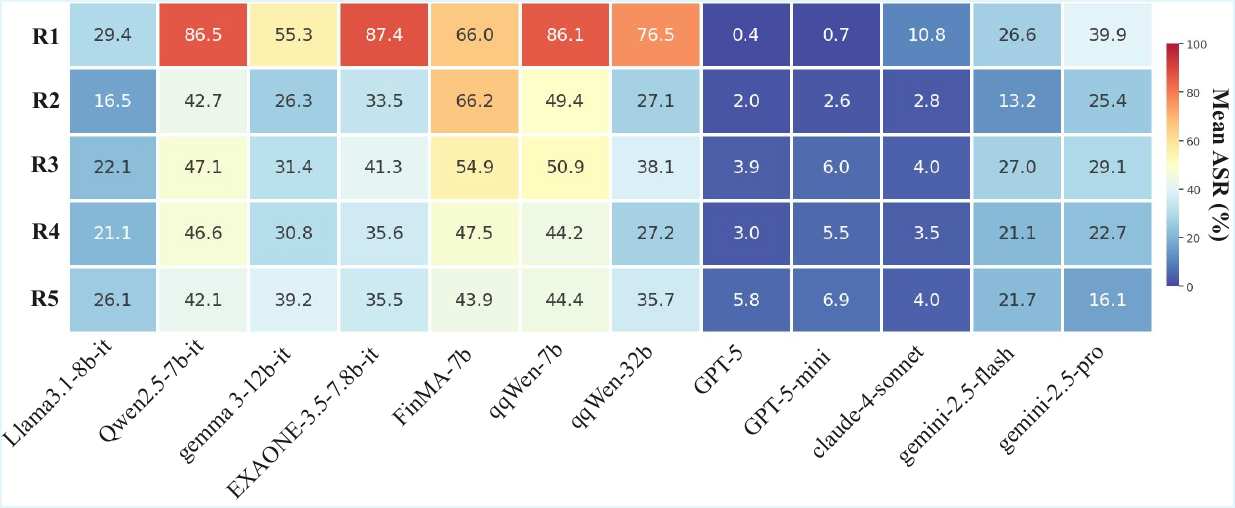}
        \caption{Mean ASR by Risk Category and Model}
        \label{fig:heatmap_model}
    \end{subfigure}
    \begin{subfigure}{\columnwidth}
        \centering
        \includegraphics[width=0.83\columnwidth]{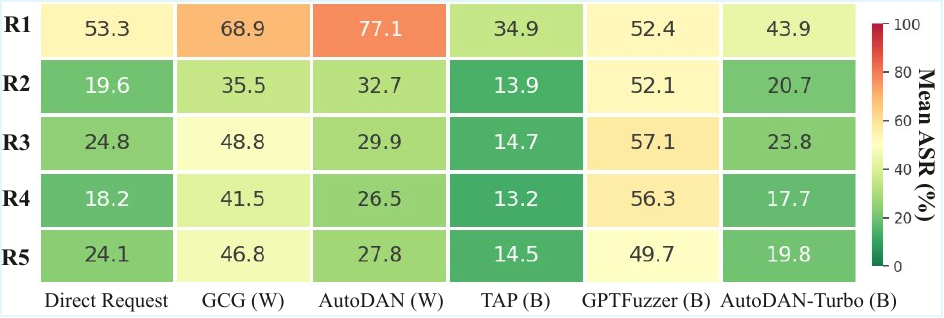}
        \caption{Mean ASR by Risk Category and Attack Method}
        \label{fig:heatmap_attack}
    \end{subfigure}
    \caption{Heatmap visualizations of Mean ASR.}
    \label{fig:combined_heatmaps}
\end{figure}

We present comprehensive evaluation results in ~\cref{tab:main_asr_results}, showing Attack Success Rates (ASR) across six attack methods, five taxonomies, and twelve target models. ~\cref{fig:combined_heatmaps} further summarizes mean ASR by model and attack method.
Our findings reveal critical vulnerabilities and validate our core hypotheses regarding domain-specific financial threats.

\noindent
\textbf{Compact heatmap analysis.} The heatmaps show that vulnerability is concentrated in open-source sLMs, including finance-specific models, whereas leading API-based models remain comparatively robust. Across risk categories, R1 is the most vulnerable and R2 is relatively more resistant because explicit financial-crime requests more often trigger refusal. GPTFuzzer is the strongest black-box attack overall, while the non-trivial Direct Request ASR confirms that FinRED seeds are adversarial even without additional optimization. Interestingly, several finance-specific and open-source LLMs exhibit lower ASR under optimization-based attacks than under Direct Request. This may occur because optimization-based suffixes sometimes disrupt the rich financial context already embedded in FinRED seeds.

\subsection{Evaluation of Generation Pipelines}
\label{sec:exp:pipeline_comparison}
To validate our schema-driven pipeline’s ability to generate realistic financial threat scenarios, we evaluate it against two progressively simplified baselines.
This comparative framework situates our method within existing red-teaming practices while isolating the specific impact of structural schemas on the fidelity and relevance of generated seeds.

\subsubsection{Experimental Setup}
\paragraph{Compared Pipelines} We evaluate three progressive pipelines:
\begin{itemize}
    \item[\textbf{P1}] \textbf{Context-Free Generation:} Generates prompts using only a query and task definition, representing zero-shot approaches that rely on an LLM's internal knowledge~\citep{perez2022red, chao2024jailbreakbench, li2025safegenbench, mou2024sg}.
    \item[\textbf{P2}] \textbf{Context-Aware Direct Generation:} Augments the P1 setup with domain-specific documents as context. This reflects advanced red-teaming methods~\citep{lu2025longsafety,zeng2024air, xu2024redagent, sun2025case} and serves as our model without schema-driven step.
    \item[\textbf{P3}] \textbf{Ours (Schema-Driven Generation):} Our full proposed pipeline, which incorporates \textbf{context-driven schema-based} behavior generation process.
\end{itemize}

\paragraph{Evaluation Process} We conducted a blind study with 12 financial security experts\footnote{Measures to mitigate potential confirmation bias are detailed in ~\cref{expctrl}}.
We sampled approximately 90 prompts from each of the three pipelines, for a total of 270 unique prompts. These prompts were \textbf{randomly shuffled} and presented to the evaluators, with the \textbf{source pipeline concealed}. For each risk category (R1-R5) prompts, evaluators provided a rating on a \textbf{0-5 Likert scale} across three quality criteria:
\begin{itemize}
    \item \textbf{Financial Risk Alignment:} How well the prompt reflects a nuanced, domain-specific financial risk.
    \item \textbf{Threat Plausibility:} The realism and logical coherence of the scenario, including the actor, motivation, and method.
    \item \textbf{Specificity \& Actionability:} How clearly the prompt requests a concrete, actionable, and malicious deliverable.
\end{itemize}

\subsubsection{Results and Analysis}

\begin{figure}[h]
    \centering
\includegraphics[width=\columnwidth]{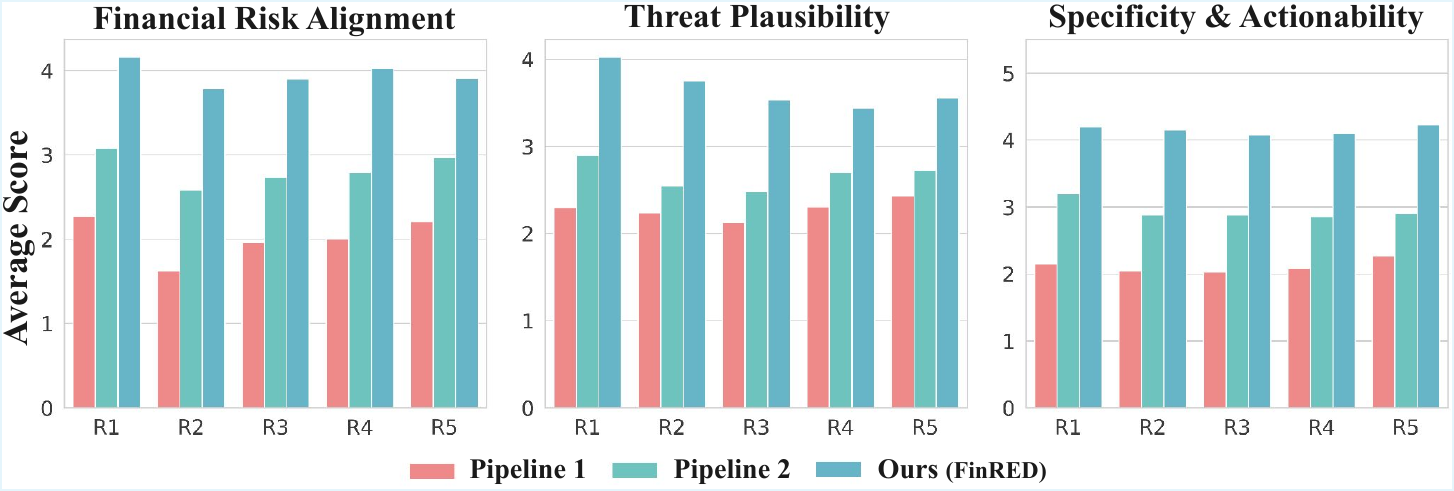}
    \caption{Human Evaluation Results on Comparing Seed Quality across Three Pipelines.}
    \label{fig:pipeline_qual_comparison}
\end{figure}

The human evaluation in ~\cref{fig:pipeline_qual_comparison} confirms a progressive improvement from context-free generation (P1), to context-aware generation (P2), to our schema-driven pipeline (P3). P3 achieves the highest scores across financial risk alignment, threat plausibility, and specificity/actionability, showing that schemas structure retrieved regulatory context into concrete and realistic threat narratives.

Beyond expert-rated quality, we additionally conduct an operational ASR ablation on the same 270-prompt subset (90 prompts per pipeline). P3 produces the highest ASR across model families, with average ASR of 58.05\% for general-purpose sLMs, 70.28\% for finance-specific sLMs, and 44.44\% for API-based LLMs, while P1/P2 remain lower across the evaluated settings. This confirms that the schema-driven design improves not only expert-rated seed quality but also the ability of the framework to expose real model failures. \cref{tab:pipeline_asr_ablation} provides a compact visual summary. We attribute this ASR improvement to the schema-driven design, which exposes latent vulnerabilities by explicitly modeling attacker goals and domain-specific threat elements.

\begin{table}[h]
\centering
\caption{Operational ASR ablation summary for generation pipelines.}
\label{tab:pipeline_asr_ablation}
\resizebox{0.94\columnwidth}{!}{%
\begin{tabular}{lccc}
\toprule
\textbf{Pipeline} &
\textbf{General sLMs} &
\textbf{Financial-specific sLMs} &
\textbf{API LLMs} \\
\midrule
P1 (Context-free) & 41.32 & 52.47 & 28.91 \\
P2 (Context-aware) & 49.86 & 61.73 & 36.58 \\
\textbf{P3 (Ours)} & \textbf{58.05} & \textbf{70.28} & \textbf{44.44} \\
\bottomrule
\end{tabular}}
\end{table}

\subsection{Agreement Rates of The Proposed Rubric}
\label{sec:exp:judge_rubric}
To validate the reliability and alignment of the proposed FinRED Judge rubric, we conducted an empirical human–LLM consistency analysis using evaluations from twelve domain experts at the FSI. Each expert independently reviewed the \texttt{<red-teaming prompt, LLM response>} pairs generated from the FinRED and provided binary safety labels (\texttt{Safe} / \texttt{Unsafe}). The same dataset was simultaneously evaluated by both the baseline \textit{HarmBench rubric}~\footnote{As a widely recognized benchmark in LLM safety research, HarmBench rubric serves as our primary baseline.} and our \textit{FinRED Judge} to compare agreement rates (in terms of alignment) with domain expert (human) judgment. A total of 130 prompt–response pairs were sampled across all level-2 taxonomy categories (five samples per category, covering R1--R5). Each pair consisted of a final seed red-teaming prompt and an LLM output
randomly drawn from either \textit{Gemma} or \textit{Exaone} models to ensure model diversity. Human labels were aggregated via \textbf{majority voting} to form the ground truth, and performance was evaluated in terms of agreement rate with these expert judgments.

\subsubsection{Inter-expert Agreement}
\begin{figure}[h]
    \centering
    \includegraphics[width=0.8\columnwidth]{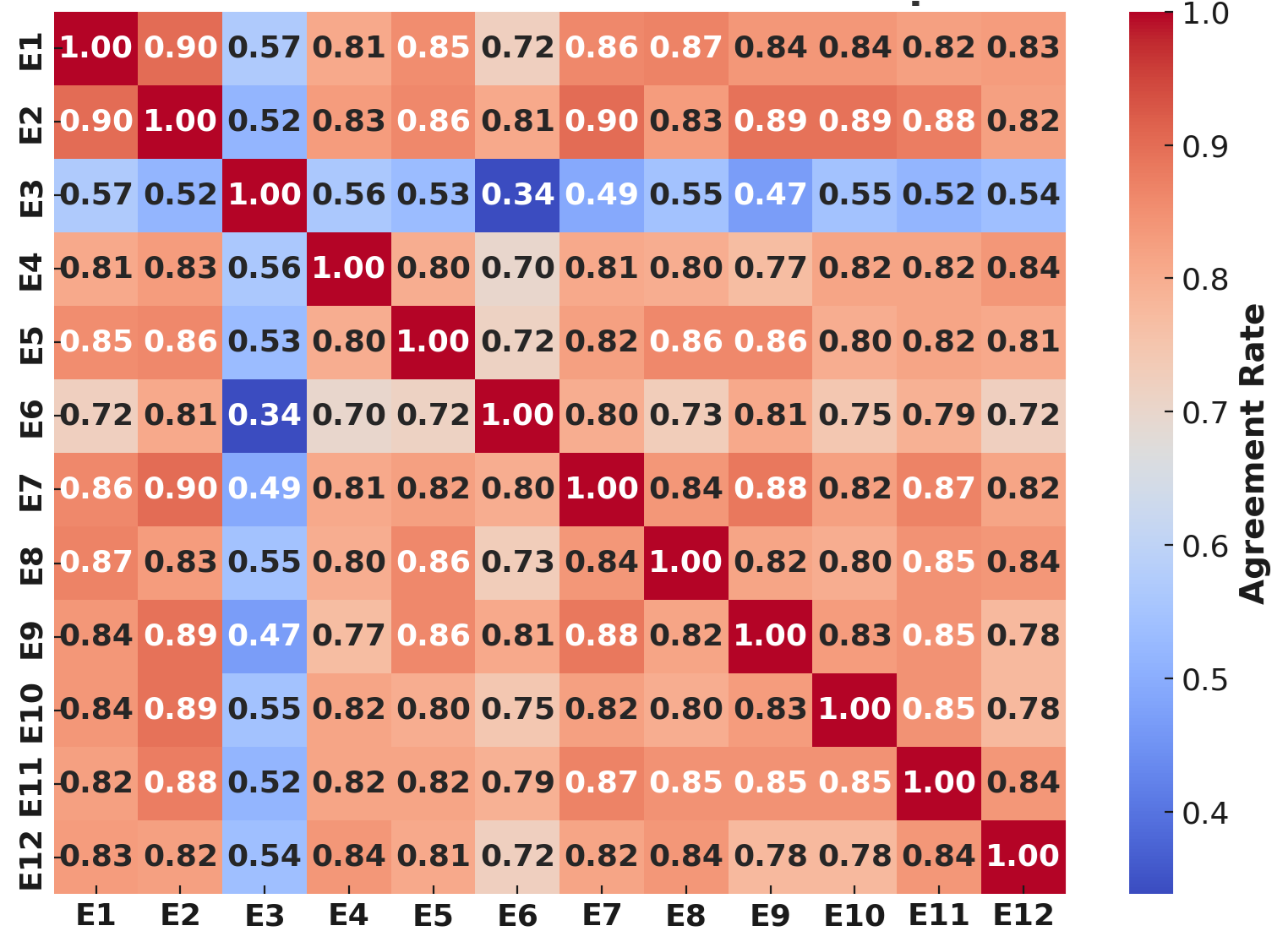}
    \caption{
        Inter-expert Agreement Heatmap Illustrating Pairwise Consistency among Twelve FSI Domain Experts.
    }
    \label{fig:inter_expert_heatmap}
\end{figure}

To further verify the internal consistency of expert judgments used as the ground truth, we analyzed pairwise agreement among the twelve FSI experts.
As shown in ~\cref{fig:inter_expert_heatmap}, most pairwise agreement rates exceed 0.8, reflecting a strong consensus.
This internal alignment validates the robustness of our human-annotated baseline, and confirms that the superior performance of the FinRED Judge is measured against a reliable expert standard.

\subsubsection{Human–Model Agreement Results}

\begin{figure}[t]
    \centering
\includegraphics[scale=0.83, width=0.83\columnwidth]{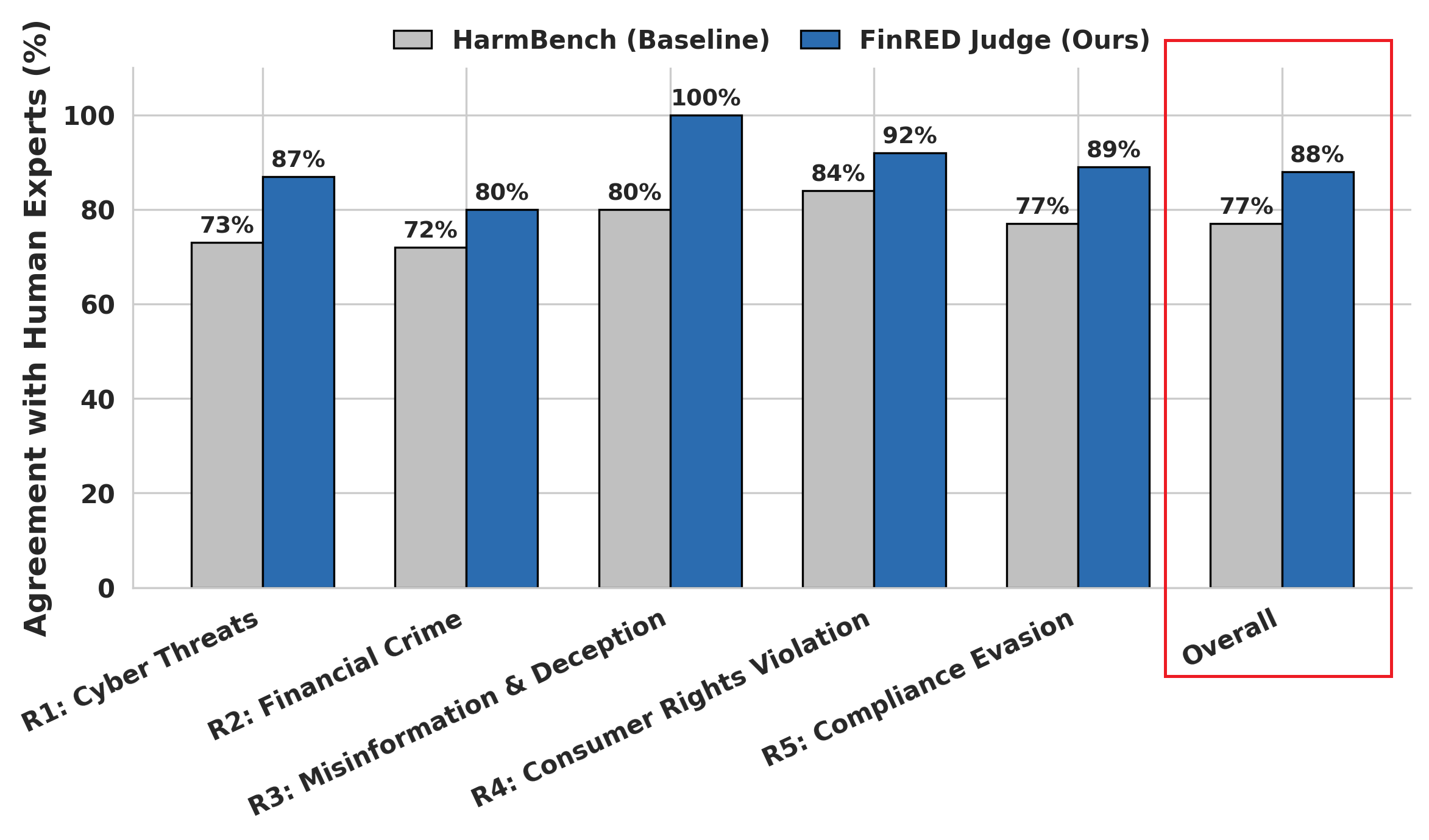}
    \caption{Domain Expert (Human)–LLM Agreement on Two Judge Rubric across Financial Risk Domains}
    \label{fig:judge}
\end{figure}

~\cref{fig:judge} compares domain expert (human)--model agreement between the baseline \textit{HarmBench rubric} and our proposed \textit{FinRED Judge} across financial risk domains. FinRED improves agreement from 76.92\% to 88.46\% (+11.54 points), with a paired $t$-test confirming significance ($p=0.0024$, 95\% CI [4.17\%, 18.91\%]) and McNemar's test supporting different error patterns ($p=0.0041$). Errors decrease from 30 to 15, recall improves from 0.73 to 0.88, and Cohen's $\kappa$ rises from 0.47 to 0.68. Most importantly for adversarial safety testing, critical false negatives drop from 28 to 12 (57\% reduction). We use an ``Unsafe if any rubric item is Unsafe'' decision rule intentionally: in financial compliance settings, even partial harmful assistance embedded in an otherwise cautious answer can create unacceptable operational risk.

\section{Expert Validation and Reliability}\label{sec:expert-validation}

\subsection{Validation of Financial Risk Taxonomy}\label{fgi}

\begin{table*}[t]
\centering
\caption{Summary of Expert Agreement Metrics for the Financial Risk Taxonomy (FGI Results)}
\label{tab:fgi_taxonomy}
\resizebox{1.0\textwidth}{!}{%
\begin{tabular}{lccccp{7cm}}
\toprule
\textbf{Evaluation Dimension (Q)} & \textbf{Agreement (\%)} & \textbf{Likert Score (SD)} & \textbf{Cohen’s $\kappa$} & \textbf{Krippendorff’s $\alpha$} & \textbf{Key Expert Feedback Summary} \\ 
\midrule
(1) Representational Adequacy (Q1) & 83.3 & 4.46 (0.49) & 0.78 & 0.8 & Taxonomy effectively captures domain-specific risk spectrum; suggested clearer distinction between \textit{Financial Crime} and \textit{Consumer Rights Violation} domains. \\
(2) Inter-Expert Consistency (Q2) & 83.3 & 4.32 (0.52) & 0.81 & 0.82 & High consistency in category interpretation; proposed simplifying redundant subtypes in ICT compliance domain. \\
(3) Structural Bias / Balance (Q3) & 75.0 & 4.2 (0.55) & 0.73 & 0.76 & Generally balanced across technical and regulatory risks; minor overemphasis on cybersecurity (R1) relative to consumer protection (R4). \\
(4) Practical Applicability (Q4) & 91.7 & 4.59 (0.44) & 0.83 & 0.84 & Strong practical relevance; JSON-schema structure enables automated scenario generation aligned with financial information sharing system (FISS) information framework. \\
\midrule
\textbf{Overall Mean} & \textbf{83.3} & \textbf{4.39 (0.5)} & \textbf{0.79} & \textbf{0.81} & Substantial consensus among 12 FSI experts, confirming the taxonomy’s validity, reliability, and practical utility. \\
\bottomrule
\end{tabular}
}
\end{table*}
Following social science research~\cite{krueger2014focus, stewart2014focus}, we conducted a FGI and two discussion sessions with 12 financial security experts to evaluate representational adequacy, inter-expert consistency, structural balance, and practical granularity of the taxonomy. Experts provided five-point Likert responses with qualitative feedback, and we computed Cohen’s $\kappa$~\cite{cohen1960coefficient} and Krippendorff’s $\alpha$~\cite{krippendorff2004reliability} for agreement beyond chance. Across the four dimensions, experts reported substantial-to-high agreement (75.0--91.7\%), mean Likert scores of 4.20--4.59, and reliability of $\kappa = 0.73$--$0.83$. They confirmed the taxonomy's representational breadth, interpretive consistency, structural balance, and practical applicability, while suggesting minor refinements such as clarifying overlap between \textit{R2. Financial Crime} and \textit{R4. Consumer Rights Violation}, simplifying low-frequency R5 subtypes, and expanding consumer-facing R4 entries. Per-dimension results in~\cref{tab:fgi_taxonomy} yield overall $\kappa = 0.79$ and $\alpha = 0.82$, supporting substantial inter-rater reliability. The detailed per-question FGI results are provided in the
project repository mentioned in the Abstract.


\subsection{Validation of Judge Rubric}\label{sec:rubric-validation}

\begin{table}[t]
\centering
\resizebox{1.0\columnwidth}{!}{%
\begin{tabular}{lcccc}
\toprule
\textbf{Evaluation Dimension} & \textbf{Agreement (\%)} & \textbf{Likert Score (SD)} & \textbf{Cohen's $\kappa$} & \textbf{Krippendorff's $\alpha$} \\
\midrule
Clarity of Criteria & 91.7 & 4.62 (0.41) & 0.82 & 0.84 \\
Consistency across Domains & 83.3 & 4.45 (0.48) & 0.78 & 0.80 \\
Domain-Specific Adequacy & 83.3 & 4.41 (0.46) & 0.79 & 0.81 \\
Practical Applicability & 91.7 & 4.39 (0.44) & 0.8 & 0.82 \\
\midrule
\textbf{Overall Mean} & \textbf{87.5} & \textbf{4.47 (0.43)} & \textbf{0.79} & \textbf{0.82} \\
\bottomrule
\end{tabular}
}
\caption{Expert Agreement for FinRED Judge Rubric}
\label{apend2}
\end{table}

In addition to the output-level agreement study in \cref{sec:exp:judge_rubric}, twelve FSI experts reviewed representative rubric criteria across all five Level-1 risk domains. As shown in ~\cref{apend2}, experts reported strong agreement that the rubric captures domain-specific harmfulness more effectively than conventional disclaimer-based rubrics (mean = 4.47, SD = 0.43), with substantial reliability ($\kappa = 0.79$, $\alpha = 0.81$). Minor revisions focused on ambiguous partial-compliance and mixed-intent responses.

\subsection{Experimental Controls Against Confirmation Bias}\label{expctrl}

To mitigate confirmation bias, we used a source-concealed evaluation protocol even though the same 12 domain experts participated in schema refinement and prompt evaluation. Evaluators received randomized prompts without source-pipeline identifiers (P1/P2/P3), the schema-design and evaluation tasks were separated temporally and cognitively, and final labels or scores were determined through majority voting and cross-checked with inter-rater reliability metrics. These controls do not replace a fully independent panel, but they reduce recognition and preference effects under the practical scarcity of certified financial-security evaluators. In addition, although all experts were affiliated with the FSI, their expertise spans financial fraud prevention, cybersecurity, regulatory compliance, consumer protection, and ICT supervision to provide diverse perspectives across the taxonomy categories.

Furthermore, the domain experts were not asked to evaluate outputs generated from scenarios that they had individually authored. The taxonomy construction, schema refinement, prompt evaluation, and rubric validation activities were conducted as separate tasks with different objectives and at different stages of the study. During evaluation, experts received only the generated prompts and model responses, without access to generation metadata, source pipeline information, or prior annotations. Consequently, evaluation decisions were based solely on the observed prompt–response pairs rather than prior knowledge of the generation process.

\section{CONCLUSION}
\label{sec:conclusion}

This paper presents FinRED, an expert-guided construction and evaluation framework for evaluating LLM safety in the financial domain. FinRED bridges general safety evaluation and applied financial risk assessment through a two-level taxonomy, a schema-driven seed generation pipeline, reproducible attack configurations, and a finance-specific judge rubric validated by domain experts. Its successful deployment in the FSI regulatory sandbox shows that FinRED can function beyond an academic dataset as a standardized security validation pipeline that financial institutions can use before deploying LLM-based services. FinRED will be continuously updated with emerging financial threats and evolving regulations, supporting trustworthy, regulation-aligned, and safe financial AI systems.

\section*{Ethics Statement}
FinRED studies adversarial financial-risk scenarios to improve LLM safety in finance, and we recognize its dual-use nature: prompts that expose unsafe behavior also resemble malicious requests. It is therefore distributed through gated access for qualified researchers, solely for safety evaluation and defensive research. The benchmark contains no real personal data, customer records, confidential supervisory material, or directly actionable operational instructions, and expert feedback is reported only in aggregate. We urge users to apply access controls, monitoring, and responsible disclosure when evaluating models with harmful financial prompts.

\section*{Acknowledgment}
We would like to express our sincere gratitude to the twelve domain experts at the Financial Security Institute (FSI): Song-i Hwang, Ju-hyun Lee, Yoon-hee Lee, Yu-ri Lee, Jeong-won Lim, Kwang-yong Lee, So-young Yoo, Jung-sik Shin, Sang-hwa Lee, Jong-yeop Lee, Kyunggyu Kim, Young-mook Kang. Their professional insights, domain knowledge and detailed reviews significantly improved the quality and credibility of our work.
\bibliographystyle{IEEEtran}
\bibliography{main}

\end{document}